# Real-Time Visualization of the Spatiotemporal Dynamics of 3D Solitons

**XinGe Liu,† Chaoyang Geng,† Yunhan Yu, Lixia Xi, Xiaoguang Zhang, Xiaosheng Xiao***
State Key Laboratory of Information Photonics and Optical Communications, School of Electronic Engineering, Beijing University of Posts and Telecommunications, Beijing 100876, China.
†These authors contributed equally to this work. *Corresponding author (xsxiao@bupt.edu.cn)

**ABSTRACT**

Three-dimensional (3D) optical solitons bear far stronger relevance to multi-dimensional nonlinear dynamics prevalent in complex physical, chemical and biological systems than conventional 1D solitons, and they support far richer and more intricate phenomena arising from their spatiotemporal degrees of freedom. However, real-time recording of 3D soliton evolution with simultaneous spatiotemporal resolution remains a critical challenging. Here, we demonstrate long-term real-time visualization of 3D soliton dynamics with spatiotemporal resolution using high-speed photodetectors combined with joint space- and time-division multiplexing. We visually capture complex transient behaviors of 3D solitons in a multimode fiber laser and, by integrating our method with the time-stretch technique, simultaneously record pulse-resolved beam and spectral evolutions. We observe that during the birth of 3D solitons, the highly multimode beam stabilizes for a substantial interval prior to spectral broadening, indicating that a large number of transverse modes have already locked before longitudinal mode proliferation. These findings highlight the critical importance of real-time spatiotemporal visualization for advancing ultrafast multimode laser design and delivering new insights into high-dimensional nonlinear dynamics.

## I. Introduction

Nonlinear dynamics are prevalent across scientific disciplines, including physics, chemistry, biology, and even sociology [1]. Laser cavities serve as an ideal experimental platform for exploring nonlinear dynamics, advancing fundamental understanding of complex systems and providing a theoretical foundation for technological development. Solitons—self-sustaining, particle-like wave packets—are widely observed across diverse nonlinear systems. Extensive studies have uncovered rich soliton dynamics in passively mode-locked lasers, including chaos, soliton explosions, and soliton molecules [2]. Real-time observation of these phenomena has been enabled by techniques such as time lens [3], spectral shearing interferometry [4], streak camera spectroscopy [5], and most notably, the time-stretch dispersive Fourier transformation (DFT) [6]. By mapping single-shot spectra roundtrip by roundtrip using highly dispersive media, DFT has revolutionized real-time studies of soliton dynamics. Over the past decade, such techniques have dramatically advanced nonlinear dynamics research, revealing transient dissipative soliton dynamics [7], soliton explosion [8], soliton molecule evolution [9,10], soliton breathing [11], and especially the formation of diverse solitons (i.e., the buildup process of mode locking) [12-15].

In conventional mode-locked lasers, which usually operate in a single transverse mode, solitons are effectively one-dimensional (1D), with fixed transverse spatial distribution. The recent achievement of spatiotemporal mode-locking (STML)—simultaneous locking of multiple transverse and longitudinal modes—in multimode fiber lasers [16] introduced transverse spatial degrees of freedom ($x$, $y$), enabling 3D solitons with far richer nonlinear dynamics. These systems greatly expand the role of lasers as platforms for nonlinear science research, offering new perspectives on high-dimensional, interdisciplinary problems such as those in fluid dynamics and Bose–Einstein condensates [17]. As an ideal testbed for high-dimensional nonlinear dynamics, spatiotemporal mode-locked lasers have attracted significant interest [18-24], revealing phenomena including multimode soliton molecules, rogue wave, period-doubling bifurcation and chaos. These observations indicate the existence of extraordinarily complex and rich nonlinear phenomena in multimode cavities. Yet few findings have fundamentally transcended the framework of 1D soliton dynamics, largely due to the lack of spatiotemporally resolved detection tools.

From the perspective of laser science, spatiotemporal mode-locked lasers support higher pulse energies and carry richer spatial information than single-mode systems, making them highly valuable for practical applications. However, the fundamental mechanisms and dynamical processes of STML remain poorly understood. Despite progress in understanding STML mechanisms [25–27], the buildup processes by which transverse and longitudinal modes lock remain unclear. Investigating the coupled evolution of transverse and longitudinal modes inside the cavity—especially the STML buildup process(i.e., the self-organization of 3D solitons)—will advance STML technology, including the rational design of laser architectures and the exploration of novel lasing regimes with superior performance or unique properties.

Real-time spatiotemporal visualization of 3D soliton dynamics is therefore essential for advancing both multimode laser physics and high-dimensional nonlinear systems, yet it poses considerable challenges: dynamic can evolve over milliseconds timescales (~$10^5$ roundtrips), with soliton characteristics changing every roundtrip (~10 ns), particularly in the transverse ($x$,$y$) profile due to the variations in spatial modal composition. Conventional image sensors lack sufficient frame rates. Compressed ultrafast photography based on streak cameras [28] captures only limited consecutive frames, unsuitable for long-duration soliton evolution. Multi-channel time-stretch DFT [29–31] enables real-time observation but suffers from severely limited spatial resolution because DFT disperses pulses into low-peak-power nanosecond waveforms, requiring substantial energy per channel for adequate signal-to-noise ratio. In addition, the stretched pulse duration, comparable to the roundtrip time, strictly limits the number of time-division multiplexed channels. Recently developed spatiotemporal DFT shows promise for resolving the spectral evolution of individual spatial modes [32], but this method is restricted to few-mode fiber cavities due to strong mode mixing, coupling, and significant higher-order mode loss in long dispersive fibers.

Here, we report long-term, real-time visualization of 3D soliton dynamics with spatiotemporal resolution, using array-based spatial sampling with time-division multiplexing (ASTM). Analogous to how DFT revolutionized 1D soliton studies by enabling single-shot spectral measurements, ASTM captures single-shot beam profiles roundtrip by roundtrip. By further combining ASTM with DFT, we simultaneously record pulse-resolved beam and spectral evolutions in a multimode cavity, which directly reflect transverse and longitudinal mode dynamics, respectively. Our observations uncover rich 3D soliton dynamics and provide design guidance for spatiotemporal mode-locked lasers, and our method can be extended to other ultrafast optical systems involving transverse spatial degrees of freedom.

## II. Results

### 1. Principle of Visualizing 3D Soliton Dynamics

Figure 1 illustrates the principle of ASTM, which was preliminarily introduced in our conference paper [33]. For an ultrafast 3D ($x$, $y$, $\tau$) optical soliton, we sample the full transverse ($x$, $y$) beam profile using an array of samplers such as optical fibers, where each sampler records local pulse energy, then the beam profile can be reconstructed from these measurements. For a pulse train with tens-of-nanoseconds separation, single-shot profiling requires rapid, parallel readout. Since each sampler outputs an ultrashort pulse, we use time-division multiplexing—setting different lengths for $n$ sampling paths to create ~1 ns temporal separation between sampled pulses, which are then combined—to distinguish samplers by arrival time with a single high-speed photodetector. $N$ photodetectors and corresponding oscilloscope channels measure $n\times N$ spatial points, enabling single-shot beam profiling with relatively high transverse spatial resolution within tens of nanoseconds. In our experiments, $n = 12$, $N = 3$, forming a 6×6 array of 36 sampling points.

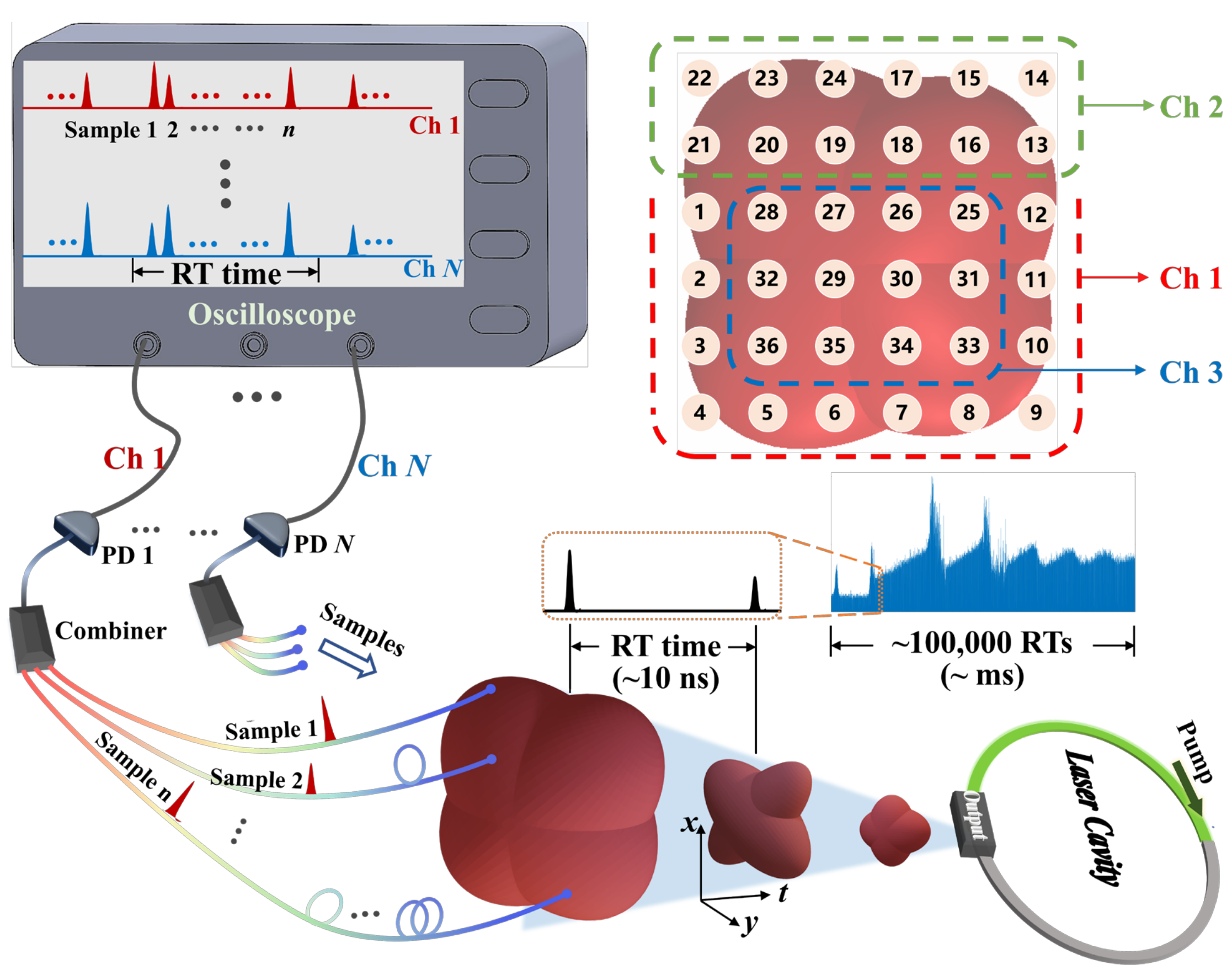

**Fig. 1 Schematic of the array-based spatial sampling with time-division multiplexing (ASTM) technique.** The transverse ($x$, $y$) field of each soliton is sampled at $N$×$n$ points, divided into $N$ channels. Each channel combines $n$ samplers and is detected by a high-speed photodetector and real-time oscilloscope. The beam profile of each soliton is reconstructed from the intensity distribution across the $N$×$n$ points. Inset (right middle): Direct measurement of soliton evolution in the laser cavity using a single photodetector and oscilloscope. The evolution process can span up to $10^5$ round trips (~ms), with a roundtrip time of ~10 ns. Inset (top right): The 3×12 sampling array used in this work. RT: roundtrip.

We applied ASTM to record the real-time dynamics of a spatiotemporal mode-locked multimode fiber laser whose output fiber supports ~100 modes with a ~18 ns roundtrip time. Detailed experimental setup and validation are provided in Methods and Supplementary Sections S1–S2. Below, we present rich observations including transient behavior of soliton evolution and single/dual soliton formation, revealing the complexity of spatiotemporal nonlinear dynamics of 3D solitons (additional dynamics including soliton transitions are shown in Supplementary Information).

## 2. Transient behavior of 3D Soliton Evolution

A representative transient evolution of 3D soliton is shown in Fig. 2. Direct detection at a single spatial point using a fast photodetector (Fig. 2a) —which closely matches the soliton energy evolution (Fig. 2b)— suggests that the system appears stable apart from six brief intensity increases followed by recovery. However, ASTM reveals critical information invisible to conventional measurements. Using intensity distributions from all 36 samplers, we reconstruct pulse-resolved transverse beam profiles (insets in Fig. 2b). The results show that the 3D soliton beam profile evolves continuously not only during the energy-varying regions but also within seemingly stable energy plateaus. To quantitatively characterize the beam-profile evolution, we calculate the cross-correlation coefficient between the beam profile of each soliton and those of adjacent roundtrips (Fig. 2b). The results reveal that the beam profile changes drastically from roundtrip to roundtrip even in the stable-energy regions, whereas the degree of beam-profile variation decreases during the rising edges of the six energy bursts.

The dynamic shown in Fig. 2 are observed immediately before soliton decay. ASTM measurements indicate that the spatial coherence is poor even within the nominally stable energy regions. Interestingly, we observe a corresponding dynamical process shortly after soliton birth, as presented in Supplementary Section S3, in which the soliton exhibits a stable beam profile within two stable-energy regions, indicating high spatial coherence in both intervals.

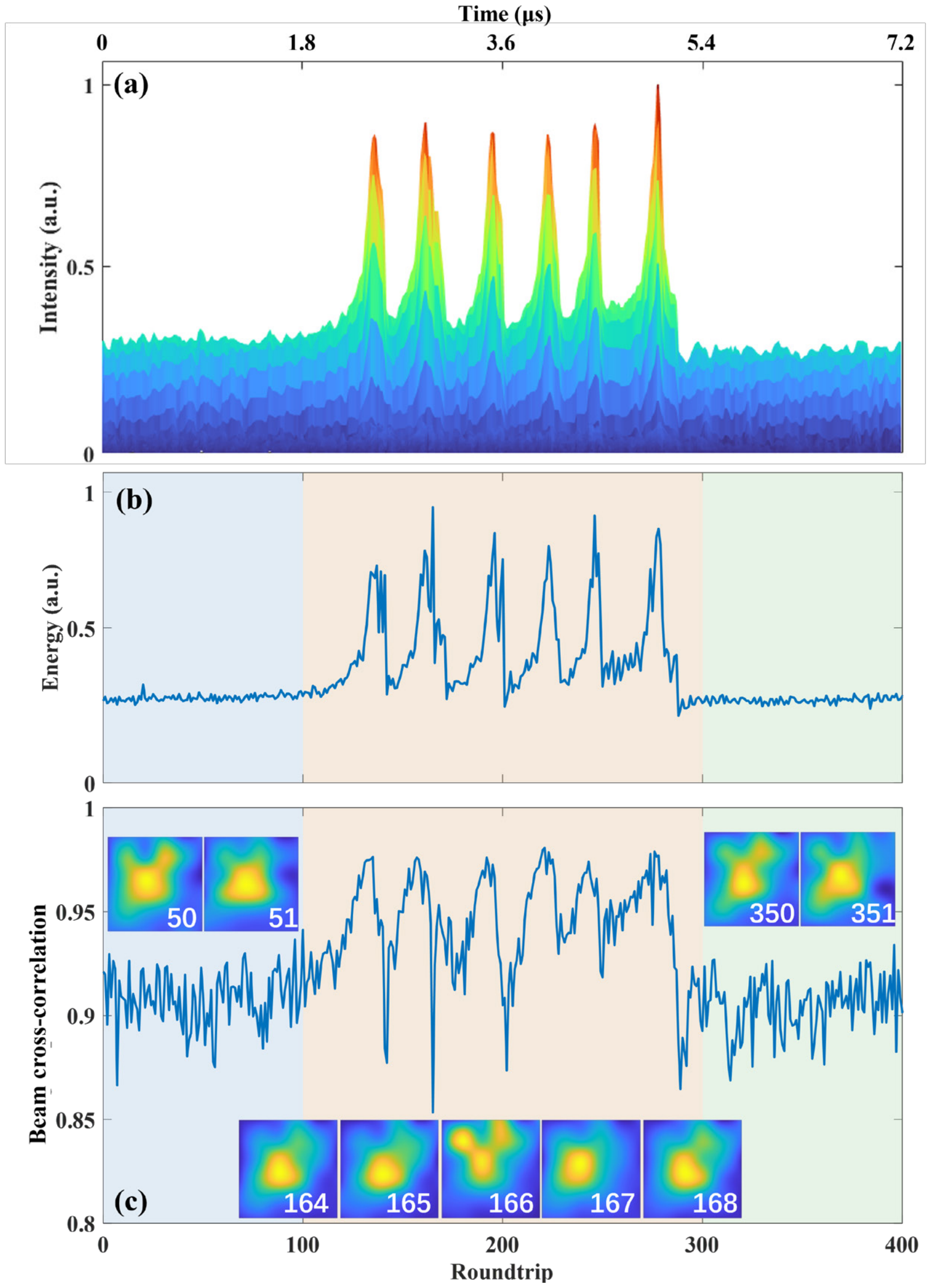

**Fig. 2 Typical transient behavior of 3D soliton evolution.** a, Dynamics measured directly by a photodetector and real-time oscilloscope at the 27$^{th}$ sampling point shown in the top-right inset of Fig. 1. b, Evolution of soliton energy, calculated as the sum of intensities measured at all 36 samplers. c, Evolution of the cross-correlation coefficient of transverse beam profiles for each 3D soliton. To reduce noise, the coefficient is averaged over the five preceding and five subsequent roundtrips. Insets (labeled with numbers) show beam profiles at representative roundtrips.

## 3. Birth of 3D solitons

ASTM not only enables visualization of 3D soliton dynamics but also can be combined with other measurement techniques to provide more complete information. In particular, integration with DFT allows simultaneous observation of pulse-resolved beam and spectral evolutions. The evolution of single-shot beam profiles reflects transverse mode dynamics in the multimode cavity, while the evolution of single-shot spectra reflects longitudinal mode dynamics, enabling simultaneous observation of the locking processes of both transverse and longitudinal modes during STML buildup.

Figure 3 shows a typical 3D soliton formation process. DFT measurement (Fig. 3b) reveals that 3D soliton formation resembles that of conventional 1D temporal solitons from the temporal and spectral domain perspectives [12]: multiple weak narrowband picosecond pulses initially coexist; a dominant pulse then emerges, with rapidly growing energy (Fig. 3d) and broadening spectrum (Fig. 3e), while other pulses decay. Concurrently with the sudden increase in pulse energy and spectral broadening driven by self-phase modulation, the beam profile also undergoes an abrupt change.

Surprisingly, before the spectral broadening of the dominant narrowband pulse (roundtrip number < 0), its beam profile is already quite stable—even under ~30% pulse-energy oscillation (Fig. 3d)—and contains a large number of higher-order modes. Since the beam profile reflects transverse mode superposition and spectrum reflects longitudinal mode superposition, our measurements indicate that during 3D soliton birth (i.e., STML buildup), transverse modes—including many higher-order modes—lock well before the number of longitudinal modes increases. As the spectrum broadens, the beam profile becomes more concentrated, indicating a reduction in higher-order modes. An animation of the pulse-resolved beam and spectral evolution is provided in Supplementary Movie S1. To our knowledge, this is the first direct experimental observation of the entire formation process of a 3D soliton including beam evolution.

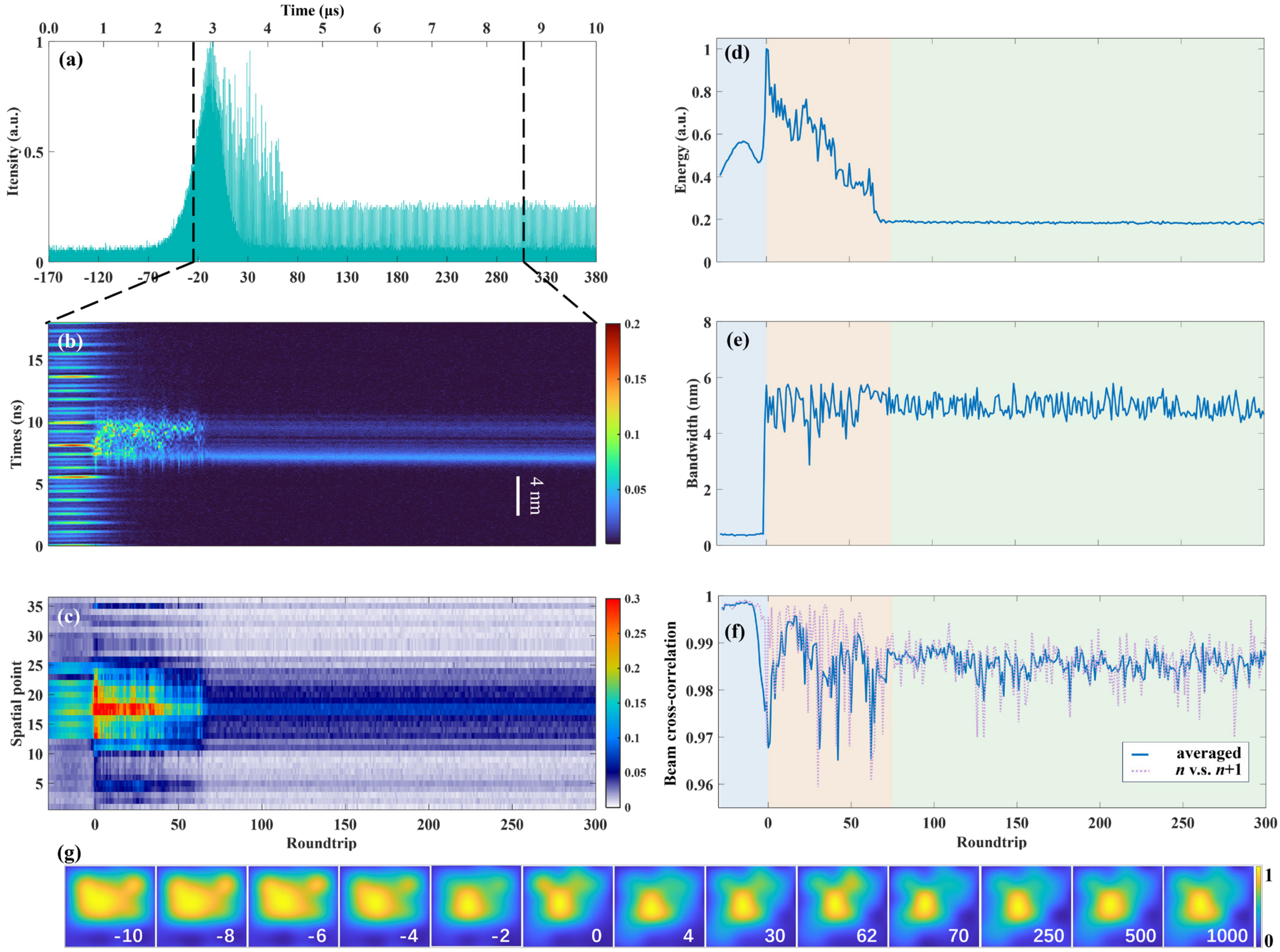

**Fig. 3 Dynamics of 3D soliton formation.** a, Raw data measured by DFT (direct output pulse evolution was not recorded simultaneously due to limited oscilloscope channels). b, Real-time evolution of single-shot spectra measured by DFT from the -28th to 300th roundtrips. c, Real-time evolution of single-shot beams measured by ASTM over the same interval as b. To intuitively show beam evolution, samplers near the beam center are plotted in the center of the *y* axis, and edge points near the edges (see Methods). d, Evolution of the dominant pulse energy, summed from all 36 samplers. e, Evolution of the 6-dB spectral width of the dominant pulse, extracted from b. f, Evolution of the cross-correlation coefficient of transverse beam profiles. The dashed line shows the coefficient between consecutive roundtrips, reflecting immediate beam changes but with higher noise; the solid line shows the average over the five preceding and five subsequent roundtrips. g, Beam profiles at representative roundtrips.

We repeated measurements under diverse cavity parameters including pump power, polarization, and intracavity coupling. Two distinct single-soliton formation processes are presented in Figs. 4a,b (visualized in Movie S2) and Supplementary Figs. S8,S9, respectively; a dual-soliton formation process is shown in Figs. 4c–e (visualized in Movie S3). Despite detailed differences, all processes—including dual-soliton

formation—follow the key conclusion from Fig. 3: the highly multimode beam stabilizes prior to spectral broadening in the early stages of soliton birth. The solitons in Fig. 4 do not reach a steady state immediately after formation. For example, the dual solitons require thousands of roundtrips to stabilize. Such long-timescale evolution can be fully captured by ASTM, with the maximum measurement duration limited only by the oscilloscope memory.

For the single soliton in Fig. 4a, after approximately the 70$^{th}$ roundtrip, the pulse energy and spectrum enter a quasi-pulsation (breathing) state with a period of ~6 roundtrips. The cross-correlation between each beam profile and that six roundtrips later (dotted curve in Fig. 4b) confirms that the beam profile also oscillates with the same period. Additional data are given in Supplementary Section S4. Figures 4c–e show the simultaneous birth of two solitons separated by 5.8 ns. As the energy suddenly rises, their beam profiles diverge significantly. However, after thousands of roundtrips, as the system stabilizes, their beam profiles gradually converge and become increasingly similar, despite large differences in their steady-state energies.

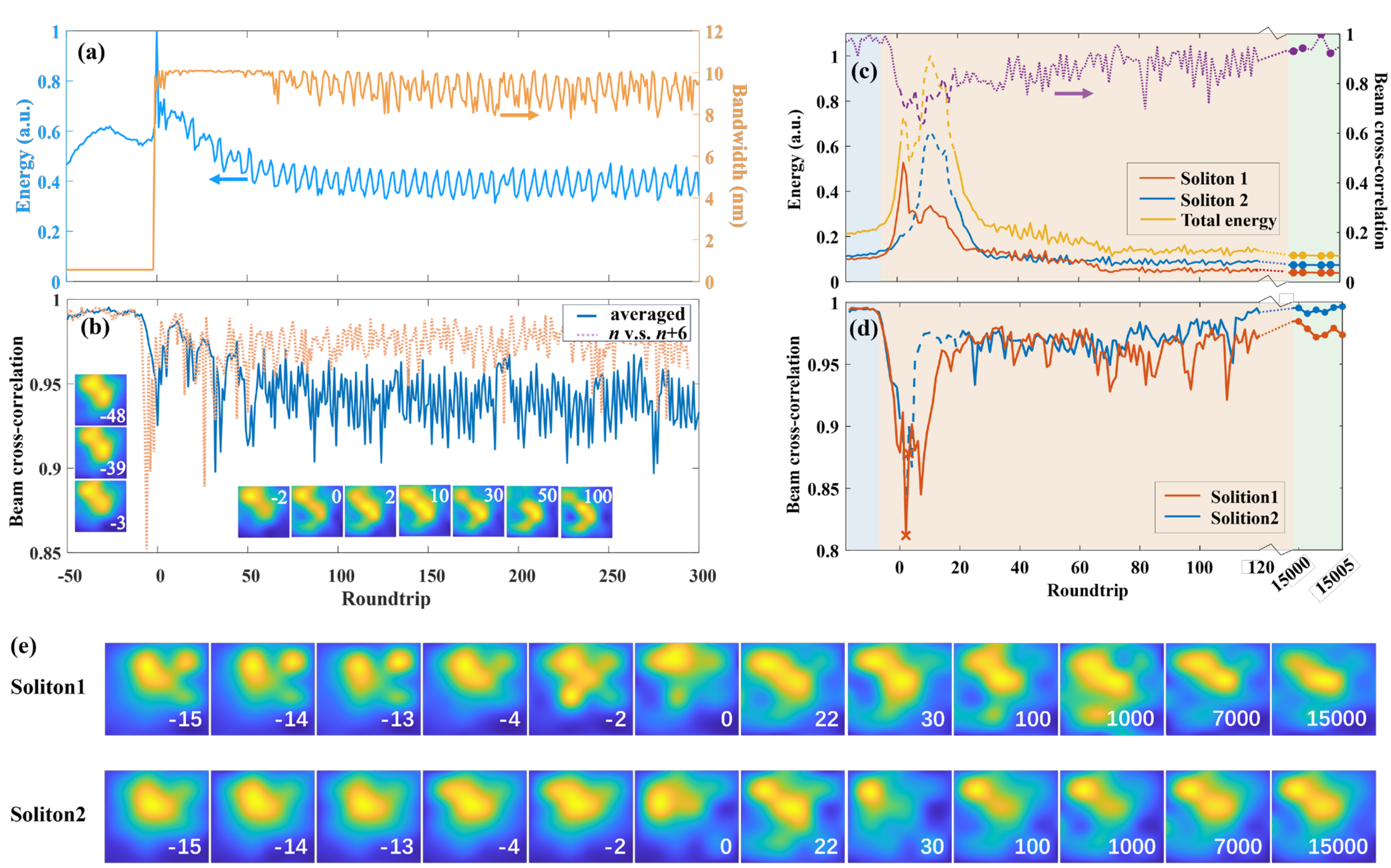

**Fig. 4 Two representative dynamics of 3D soliton formation.** a–b, Birth process of a single soliton. a, Evolution of the dominant pulse energy and 6-dB spectral width. b, Evolution of the cross-correlation coefficient of beam profiles. The blue solid curve shows the average over ±5 roundtrips; the tan dotted line shows the coefficient between each soliton and the soliton six roundtrips later. Insets show beam profiles at representative roundtrips. c–e, Birth process of dual solitons. c, Lower: Evolution of individual and total energies of the two dominant pulses. Upper: Evolution of the cross-correlation coefficient between the beam profiles of the two solitons in each roundtrip. Evolutions from –17$^{th}$ to 120$^{th}$ and 15000$^{th}$ to 15005$^{th}$ roundtrips are shown. d, Evolution of average beam-profile cross-correlation coefficients for each soliton over ±5 roundtrips. e, Beam profiles of the two solitons at representative roundtrips. In c–d, dashed segments (the 1$^{st}$–21$^{st}$ roundtrips) indicate unreliable data due to saturation of several samplers caused by excessive intensity of soliton 2.

## III. Discussion

The significance of ASTM can be appreciated by comparing it with DFT, which has revolutionized 1D soliton dynamics research. DFT enables real-time detection of single-shot spectra by mapping the spectral information into the time domain via large dispersion. Its spectral resolution, typically ~1 nm, is determined by the magnitude of the dispersion and the bandwidth of the subsequent photodetector and oscilloscope. In ASTM, the spatial-sampling array enables real-time detection of single-shot beam profiles, with spatial resolution determined by sampler density. Similar to DFT, ASTM selectively responds to pulsed signals and rejects continuous-wave background due to time-division multiplexing; thus the measured beam profiles originate exclusively from 3D solitons.

Our method has limitations that can be improved in future work. Higher spatial resolution can be achieved by increasing the number of multiplexed spatial samplers, using faster photodetectors and higher-bandwidth oscilloscopes, or by targeting longer cavities with larger roundtrip times. Currently, we measure the integrated beam profile rather than the full spatiotemporal ($x$, $y$, $\tau$) distribution. Future integration of multiple spatial-sampling DFT channels will enable more complete characterization.

In the present work, our measurements provide a qualitative rather than precise quantitative measurement of transverse intensity distribution, similar to how DFT provides approximate spectral information. Nonetheless, the 6×6 sampling array used here is fully sufficient to resolve dynamical trends and state transitions. With sufficient sampling density, ASTM can potentially enable mode-resolved dynamics by combining the reconstructed beam profiles with established numerical methods for modal decomposition, such as neural networks [34], stochastic parallel gradient descent algorithms [35], or even direct matrix operations [36].

The novel phenomena revealed by ASTM may stimulate the development of new theories. For example, our observations raise some important questions: What triggers a stable 3D soliton to abruptly restructure its modal content and switch to another stable state (Fig. S7)? What mechanism governs the long-range interaction between dual 3D solitons during their formation (Figs. 4c-e)? Beyond spatiotemporal mode-locked lasers, ASTM can be applied to a wide range of systems requiring real-time measurement of non-repetitive 3D optical pulse trains, such as multimode mesoresonators [37], single-pass propagation through multimode fibers with random coupling or noise-induced nonlinear processes, and pump-probe or dual-comb measurement systems incorporating transverse spatial information. Real-time spatiotemporal visualization will accelerate the development of ultrafast multimode lasers and deliver fundamental insights into high-dimensional nonlinear dynamics.

## Methods

**Experimental setup.**

The configuration of the multimode cavity is similar to our previous work [26]. The laser cavity consists of a 0.6-m-long multimode gain fiber and a ~2.4-m-long passive multimode fiber (with minor length variations across experiments). The gain fiber is a step-index fiber with a 20-μm core diameter supporting 6 LP modes; the passive fiber is a graded-index fiber with a 50-μm core diameter (from which 3D solitons are output) supporting approximately 100 LP modes. The laser achieves spatiotemporal mode-locking (STML) via nonlinear polarization rotation. Self-starting STML occurs with appropriate cavity alignment. The fundamental repetition frequency of the cavity is ~55.5 MHz, corresponding to a ~18 ns roundtrip time.

At the laser output, a beam splitter directs most of the optical power to the ASTM system, with the remaining small fraction serving as input to the time-stretch dispersive Fourier transform (DFT) system, enabling simultaneous measurements using both techniques. For ASTM, the beam is first magnified without distortion using a 4f system and then spatially sampled by a 6×6 square array of 36 samplers. The samplers are multimode OM4 fibers with 50-μm core diameter and numerical aperture NA = 0.2. The 36 fibers are divided into three groups. Within each group, the fibers have unique lengths differing by ~0.2 m (corresponding to a propagation time difference of ~1 ns). Three customized combiners, each combining 12 sampling fibers, feed three high-speed photodetectors (5 GHz bandwidth), whose outputs are recorded by three channels of a high-speed real-time oscilloscope with a 50 GSa/s sampling rate and 20 GHz bandwidth. For DFT measurements, light is coupled into a 6.8-km-long standard single-mode fiber using a collimator with a single-mode fiber pigtail. The output is detected by a 5-GHz-bandwidth photodetector and recorded on the fourth oscilloscope channel. Additional setup details are provided in Supplementary Section S1.

**Data processing.** The relative pulse energies are calculated as the sum of intensities measured from all 36 samplers. Beam cross-correlation coefficients are computed using the measured intensities at the 36 samplers. When unfolding the 2D beam profile into a 1D display (e.g., Fig. 3c), samplers are roughly ordered by intensity to produce a representation where higher intensities appear near the center of the *y* axis and lower intensities near the edges. The sampler indices (show in the top-right inset of Fig. 1) corresponding to $y$-axis values 1 to 36 in Fig. 3c are: [6, 1, 13, 7, 2, 3, 4, 5, 12, 18, 8, 14, 11, 17, 10, 9, 16, 15, 21, 22, 23, 29, 24, 28, 27, 20, 30, 35, 35, 33, 32, 26, 19, 25, 36, 31]. Details of beam-profile reconstruction are provided in Supplementary Section S2. Beam profiles reconstructed by ASTM show good agreement with the fundamental mode ($LP_{01}$) distribution when measuring dynamics in single-mode fibers (see Supplementary Section S2), confirming the validity of the ASTM technique. Therefore, all beam profiles presented in this work are reconstructed directly from measured data without further calibration.

**Acknowledgements**

This work was supported by National Natural Science Foundation of China (62375024) and Fundamental Research Funds for the Beijing University of Posts and Telecommunications (2025JCTP09).